\documentclass[aps,prd,preprint,groupedaddress]{revtex4}
\usepackage{graphicx}
\usepackage{amssymb}
\begin{document}

\preprint{MCTP-04-55}


\title{The Timescale for Loss of Massive Vector Hair by a Black Hole \\
and its Consequences for Proton Decay} 
\author{Andrew Pawl}
\email{apawl@umich.edu}
\affiliation{Michigan Center for Theoretical Physics, 
University of Michigan, Ann Arbor, MI, USA}

\date{\today}

\begin{abstract}
	It has long been known that matter charged under
a broken U(1) gauge symmetry collapsing to form a black hole will
radiate away the associated external (massive)
gauge field.  We show that the timescale for
the radiation of the monopole component of the field will be on the order of
the inverse Compton wavelength of the gauge boson (assuming natural units).  
Since the Compton wavelength
for a massive gauge boson is directly related to the
scale of symmetry breaking, the timescale for a black hole to lose
its gauge field ``hair'' is determined only by this scale.  The timescale
for Hawking radiation, however, is set by the mass of the black hole.  
These different dependencies mean that for any (sub-Planckian) scale
of symmetry breaking we can define a mass below which black holes
radiate quickly
enough to 
discharge themselves via the Hawking process before the gauge field
is radiated away.  This has important implications for the extrapolation
of classical black hole physics to Planck-scale virtual black holes.  
In particular, we comment on the implications for protecting protons from
gravitationally mediated decay.
\end{abstract}

\pacs{12345} 

\maketitle

\section{Introduction}

In the early 1970's Beckenstein and Teitelboim proved 
that a static
bare black hole can be endowed with no exterior classical massive
or massless scalar fields, nor with exterior classical massive
vector fields \cite{beck}.  At the same time, Price explored the
mechanism by which matter collapsing to form a black hole could
divest itself of a classical massless scalar field \cite{Prisc}.  He
concluded that the mass of the nascent black hole will set the
timescale for a massless scalar monopole field to radiate away.  

This leaves open the question of the timescale for the
loss of a massive vector field.  The question is nontrivial, since
setting the vector mass to zero results in the case of electromagnetism
(unbroken U(1) gauge symmetry).
It is well known that
an external electric monopole field can persist even in the
limit of a static black hole, where we recover the Reissner-Nordstr\"om
solution.
We are led, then, to consider whether a massive vector monopole will
decay with a timescale set by the vector boson mass or whether (as 
in the case of a scalar field) the
timescale is determined by the black hole mass.  In the former
case we recover electromagnetism in the continuous 
limit of small boson mass, while in
the latter situation a massless photon represents a discontinuous
jump from the physics of massive vector fields.

Coleman, Preskill
and Wilczek, in their work on discrete gauge symmetries and black
holes, state (without proof)
that the correct answer is the continuous one and that
the lifetime of a classical massive vector monopole field is set by the 
mass of the field itself \cite{coleman}.  In this work, we show
that their assertion is correct.  We demonstrate this
by adapting the detailed calculational
techniques used by
Price in his study of massless vector fields 
\cite{Prisc,Price} to the case of a massive vector monopole field.

This answer has important consequences for black hole
phenomenology.  Black holes are postulated to radiate via the Hawking
process \cite{hawkrad}.  When charged particles fall into a black
hole, this process can theoretically recover the ``lost'' charge
if and only if some external field generated by those charges 
remains \cite{Gibbons,Carter,Page}.  Since black holes cannot sustain
an external massive vector field forever, it has been assumed that
particles charged only under a broken U(1) gauge symmetry that fall
into a black hole result in charge non-conservation -- their charge
will not be recovered through Hawking radiation.  Now we see that
this need not be the case for all black holes.  The timescale for
Hawking radiation is related only to the mass of the black hole.
Therefore, because the external field's decay timescale and the black hole's
radiation timescale depend on parameters that are totally 
independent of one another, we 
can define a regime in which the black hole will discharge
quickly enough to respect a broken symmetry.  

This result has particular relevance to the assumptions 
made about virtual (Planck-scale) black holes and the
so-called ``spacetime foam'' (see, e.g. \cite{Hawking}).  It has been
assumed that since classical black holes do not respect a broken
symmetry, quantum processes involving black holes will also violate
such symmetries.  In this work, however, we will show that the smallest
classical black holes \emph{will} in general respect broken symmetries.
Thus, the extrapolation of conservation laws
to quantum black holes must be re-examined.

\section{Despun Field Equation for a Massive Vector Boson}

We will consider the collapse of a star containing
matter that is charged under a broken U(1) symmetry by adapting
the model defined in \cite{Prisc,Price} for the
cases of stars acting as a source of massless scalar and vector
fields to our purpose.  We will be concerned only with
the field external to the star (and later the black hole) so 
we will work in the Schwarzschild geometry with line element:
\begin{equation}
	ds^{2}=\left(1-\frac{2M}{r}\right)dt^{2} - \left(1-\frac{2M}{r}\right)^{-1}dr^{2}
-r^{2}d\theta^{2}-r^{2}\sin^{2}\theta d\phi^{2}.
\end{equation}
We will make the assumption that the charges in the collapsing star are small enough
that the vector fields never have a significant impact on the external geometry.

To analyze the behavior of the external massive vector field, it will be very convenient
to use the ``despun'' versions of the field equations, as emphasized by Price \cite{Price}.
This formalism is briefly reviewed in Appendix \ref{sec:appa}.  In the case of a purely monopole
field, the field equations reduce to:
\begin{equation}
\label{eq:fieldphi}
	-\partial_{r^{*}}^{2}\Phi_{0}+\partial_{t}^{2}\Phi_{0} - \partial_{r^{*}}
	\left[\left(1-\frac{2M}{r}\right)\frac{2}{r}\Phi_{0}\right]=-\mu^{2}
	\left(1-\frac{2M}{r}\right)\Phi_{0}
\end{equation}
where $r^{*}$ is defined by:
\begin{equation}
\label{eq:rstar}
	r^{*} = r + 2M \ln\left(\frac{r}{2M}-1\right) + C
\end{equation}
and $C$ is an arbitrary constant. It is important to 
note that in these coordinates the horizon
of the black hole will be found at $r^{*} = -\infty$.  Note also that
the quantity $\mu$ appearing in Equation (\ref{eq:fieldphi}) is the inverse
Compton wavelength of the vector boson field.  We will assume throughout
this paper that:
\begin{equation}
\label{eq:mulim}
	\mu M \ll 1.
\end{equation}
We will comment on this assumption in Section \ref{sec:consequences}.
For a monopole field, as explained in the Appendices, the despun quantity $\Phi_{0}$ 
appearing in Equation (\ref{eq:fieldphi}) is simply
the radial component of the electric field (or the equivalent for
the massive vector boson field).  As a check on the form of our
field equation, then, it is
important to notice that in the limit $r \gg 2M$ Equation (\ref{eq:fieldphi}) admits
the flat space Proca solution:
\begin{equation}
\label{eq:eform}
	\Phi_{0} = \frac{\mu e^{-\mu r}}{r} + \frac{e^{-\mu r}}{r^{2}}.
\end{equation}

To make full use of Price's results, we must make one further simplification.  
We make the change of variables $\Phi_{0} = \Psi/r$ which gives the convenient
form:
\begin{equation}
\label{eq:fieldpsi}
	- \partial_{r^{*}}^{2}\Psi + \partial_{t}^{2}\Psi = 
	\left(1-\frac{2M}{r}\right)\left[\frac{6M}{r^{3}}-\frac{2}{r^{2}}-\mu^{2}\right]\Psi.
\end{equation}

\section{Calculation of Decay Timescale}

\subsection{Reflection of the Field from the Curvature Potential}
We are now in a position to use the field equations of the previous section to 
determine how long a massive vector monopole field will persist outside of a
collapsing star.  In his analysis of scalar monopole fields, Price used both analytical
and numerical
methods to show that the timescale for decay is set by $M$, the mass
of the collapsing star \cite{Prisc}.  We will attempt to adapt both of his
methods to
the case of a massive vector field.  

We begin with the analytical approach.  Price was able to show that the
field equation for the despun piece of any radiatable massless field will
reduce to the same form exhibited by Equation (\ref{eq:fieldpsi}) -- namely,
the field equation will be a
one-dimensional wave equation with an effective potential \cite{Prisc,Price}.
In the case of our massive vector monopole field, the effective potential
is:
\begin{equation}
\label{eq:veff}
	V_{eff} =
	\left(1-\frac{2M}{r}\right)\left[\frac{2}{r^{2}}+\mu^{2}-\frac{6M}{r^{3}}\right].
\end{equation}

\begin{figure}
\includegraphics[width=\columnwidth]{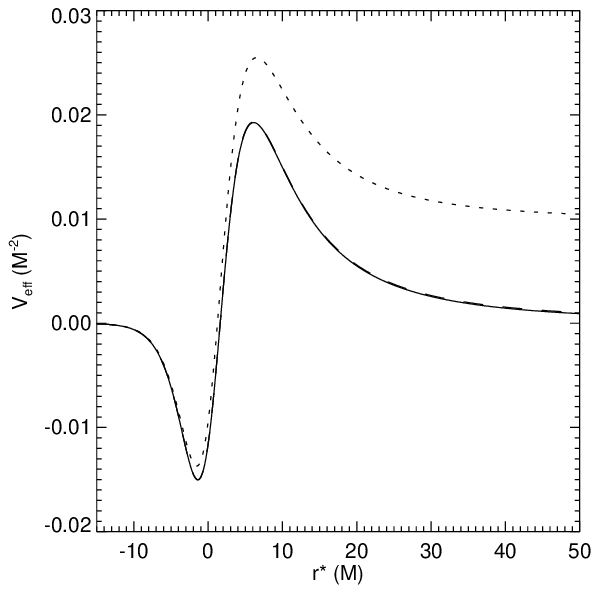}
\caption{Effective potential of Equation (\ref{eq:veff}) as a function of $r^{*}$, 
taking $C=0$ in Equation (\ref{eq:rstar}).  The potential is plotted
for three values of $\mu$.  The solid line represents $\mu=0$ (unbroken symmetry), the dotted line
$\mu=0.1$, and the dashed line $\mu=0.01$.  The line for $\mu=0.01$ is almost indistinguishable 
from the massless case.  Note that the plot for $\mu=0.1$ shows that  
the effects of $\mu$ on the potential are suppressed below $r^{*} = 3M$ ($r \sim 3.5 M$). } 
\label{fig:procapot}
\end{figure}

This effective potential plays an important role in the loss of the external
field after collapse of the star \cite{Prisc,Price,Thorne,Misner}.  As seen
in Figure \ref{fig:procapot}, the barrier is highly localized at a specific
value of $r^{*}$ (corresponding to $r \sim 3.5M$).  The precise value
of $r^{*}$ depends on the choice of the arbitrary 
constant in Equation (\ref{eq:rstar}).
Waves propagating
out from the surface of the star as it collapses beyond the location of the
barrier will have to penetrate the potential barrier if the field far outside is to remain
coupled to the sources within the star.  

As emphasized in Price's approach \cite{Prisc,Price,Thorne,Misner}, we will be
most interested in the behavior of long wavelength waves as they approach the
barrier.  This is because time dilation requires that the field on the 
stellar surface settle down to a constant value as:
\begin{equation}
	\Psi = a + b\exp\left(-\frac{t-r^{*}}{4M}\right) 
\end{equation}
where $a$ and $b$ are constants.  As the stellar surface
approaches the horizon ($t \rightarrow \infty$ and
$r^{*} \rightarrow - \infty$) the field becomes constant.  This implies an 
effective redshift of the information propagating out from the star. 
If the external field is to remain in contact with the sources falling into
the forming black hole, it can only do so via waves with wavelengths approaching
infinity.  
We can find the transmission and reflection coefficients for such waves by 
examining the solution to the wave equation (\ref{eq:fieldpsi}) in the three
distinct regions (far outside the peak of the
potential, the region where the potential
is important, and well inside the peak of the potential) 
and then matching these solutions
\cite{Thorne}.  
We will assume
\begin{equation}
\label{eq:wavesol}
 \Psi=\psi(r)\exp(-i\omega t).
\end{equation}

We begin with the solution well outside the peak of the potential.  In mathematical
terms, we want the limit $r^{*} \gg 2M$.  Note that for very large values of $r^{*}$,
Equation (\ref{eq:rstar}) implies $r \sim r^{*}$. 
In this limit, the differential equation for $\psi(r)$ becomes:
\begin{equation}
         \frac{d^{2}}{{dr^{*}}^{2}}\psi_{\rm far} + \omega^{2}\psi_{\rm far} -
        \left[\frac{2}{(r^{*})^{2}}+\mu^{2}\right]\psi_{\rm far}=0
\end{equation}
which has a solution in terms of spherical Bessel functions
of the first order \cite{Thorne,absteg}.
To find the transmission coefficient for waves propagating out from the black
hole, we must choose the linear combination which reduces to the form
$T \exp(ikr^{*})$ in the limit of very large $r^{*}$.  
This turns out to be:
\begin{equation}
\label{eq:far}
	\psi_{\rm far} = T h^{(1)}_{1}\left(\sqrt{\omega^{2}-\mu^{2}}\;r^{*}\right).
\end{equation}

Next we look at the solution in the region where the potential is important.
Recall that we are assuming $\mu M \ll 1$ so that in this region the
$\mu^{2}$ contribution is negligible.  Further, as we have
stated, we are only interested in the reflection and transmission coefficients
for long wavelength modes.  Thus, upon assuming the time dependence of Equation
(\ref{eq:wavesol}), we lose no information by taking the limit $\omega \ll M^{-1}$.
With all these simplifications,
the field equation for $\psi$ near the peak of the potential becomes::
\begin{equation}
         \frac{d^{2}}{{dr^{*}}^{2}}\psi_{\rm peak} +
        \left(1-\frac{2M}{r}\right)\left[\frac{6M}{r^{3}}-\frac{2}{r^{2}}\right]\psi_{\rm peak}=0.
\end{equation}
This equation has exact solutions.  One
solution 
is easy to guess, as this equation is exactly the same field equation we would
have found for a static, massless vector boson field.  (The limits of small
$\omega$ and small $\mu$ make this apparent.)  This indicates
that one valid solution is $\psi_{\rm peak} \propto 1/r$.
The full solution
near the peak can be written:
\begin{equation}
\label{eq:middle}
	\psi_{\rm peak} = \frac{A}{r} + B
	\left[4M^{2} + Mr +\frac{r^{2}}{3} + \frac{8M^{3}}{r}
	 \ln\left(r-2M\right)\right]
\end{equation}
where $A$ and $B$ are constants.
	
	Finally, we consider the region inside the potential.  Here, we
are in the limit of large negative values of $r^{*}$.  In this limit, the
term $(1-2M/r)$ causes the potential to vanish exponentially rapidly in
$r^{*}$.  
For this reason waves propagate freely near the horizon
in $r^{*}$ coordinates.  We can therefore adopt the solution (for
outgoing waves):
\begin{equation}
\label{eq:near}
	\psi_{\rm near} = \exp(i\omega r^{*}) + R\exp(-i\omega r^{*}).
\end{equation}	

	We now solve for the transmission and reflection coefficients
by matching Equations (\ref{eq:far}), (\ref{eq:middle}) and (\ref{eq:near}) to form
a complete solution.  We begin by matching the far zone to the region near
the potential peak.  Recall that we are interested in the limit of long
wavelength, so we can approximate the spherical Hankel
function of Equation (\ref{eq:far}) as:
\begin{equation}
	\psi_{\rm far} \sim \frac{(\omega^{2}-\mu^{2})(r^{*})^{2}T}{3} 
	- i \frac{T}{\sqrt{\omega^{2}-\mu^{2}}\;r^{*}}
\end{equation}
where we have kept the leading order real and imaginary parts.
Similarly, Equation (\ref{eq:middle}) is of the form:
\begin{equation}
	\psi_{\rm peak} \sim \frac{A}{r^{*}} + \frac{B (r^{*})^{2}}{3}.
\end{equation}
Matching, we can see that:
\begin{equation}
\label{eq:A1}
	A = -i\frac{T}{\sqrt{\omega^{2}-\mu^{2}}}
\end{equation}
and:
\begin{equation}
\label{eq:B1}
	B = T (\omega^{2}-\mu^{2}).
\end{equation}

	Next, we match the near zone to the region near the peak.
Here, for long wavelength, Equation (\ref{eq:near}) becomes:
\begin{equation}
	\psi_{\rm near} \sim 1 + R + i\omega \left(1-R\right)r^{*}.
\end{equation}
Equation (\ref{eq:middle}) has the limit:
\begin{equation}
	\psi_{\rm peak} \sim \frac{A}{2M} + 2M B r^{*}.
\end{equation}
Matching gives:
\begin{equation}
\label{eq:A2}
	A = 2M\left(1+R\right)
\end{equation}
and:
\begin{equation}
\label{eq:B2}
	B = \frac{i\omega}{2M}\left(1-R\right).
\end{equation}		

By equating our two expressions for $A$ and our two expressions
for $B$ we can arrive at two equations in the unknowns $T$ and $R$
which are solved to obtain:
\begin{equation}
	T = \frac{2i}{\frac{1}{2M\sqrt{\omega^{2}-\mu^2}}
	+\frac{2M}{\omega}\left(\omega^{2}-\mu^{2}\right)}
\end{equation}	
and
\begin{equation}
	R = \frac{1-\frac{4M^{2}}{\omega}\left(\omega^{2}-\mu^{2}\right)^{3/2}}
	{1+\frac{4M^{2}}{\omega}\left(\omega^{2}-\mu^{2}\right)^{3/2}}.
\end{equation}

\begin{figure}
\includegraphics[width=\columnwidth]{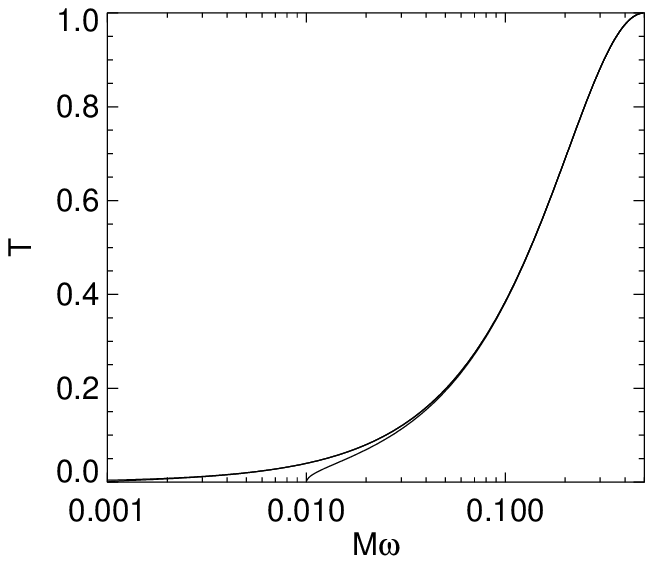}
\caption{Magnitude of the transmission coefficient ($|T|$) for outgoing massive vector waves to
propagate through the effective potential barrier as a function of 
the wave frequency $\omega$.  Plots for the
values $\mu M = 0.01$, 0.001, and 0.0001 are overlaid.  Note that the
curves are essentially indistinguishable.}
\label{fig:tloss}
\end{figure}

We are left with a paradox.  It seems that as $\omega$ approaches $\mu$,
the transmission coefficient drops to zero.  Thus, we anticipate that, as
discussed in \cite{Prisc,Price}, the external field will die away as the
information from the source (the charge of the collapsing star) is
reflected back and lost.  It is important to note, however, that the 
transmission coefficient goes to zero in a way that is essentially independent
of the precise value of $\mu$ (see Figure \ref{fig:tloss}).
 As long as $\mu$ is small compared to
$M$, we see much the same behavior for $T$.  This is a problem for 
the theory, since we find that even 
in the limit $\mu = 0$ taking $\omega \rightarrow 0$ gives $T=0$.  
This would mean that even in the case
of unbroken U(1) symmetry the external gauge field can be lost.

The loophole that saves the case of the photon (or other massless gauge
boson) is contained in the setup of the problem rather than the 
solution.  In the simultaneous limit $\mu = 0$ and
$\omega = 0$, we do not actually have
three regions worth of solutions to match to one another.  
In fact, when we take these limits, the exact solutions we derived for Equation
(\ref{eq:middle}) will be valid at \emph{any} $r$.
This is a very important point, as we shall see.

The fact that the exception for massless bosons is not obvious leads us
to be suspicious of the case of nonzero $\mu$.  The most dangerous
case is $\omega \rightarrow \mu$, since here it seems that the potential in the far zone will
strongly resemble that for the massless photon.  There is
an important difference, however.  A nonzero mass
requires
that we retain the term
$2M\mu^{2}/r$ in the potential.  Thus, for $\omega \approx \mu$, the
equation for $\psi$ in the far zone becomes:
\begin{equation}
\label{eq:omegeqmu}
         \frac{d^{2}}{{dr^{*}}^{2}}\psi_{\rm far} + \left[(\omega^{2}-\mu^{2})
        + \mu^{2}\frac{2M}{r^{*}} - \frac{2}{(r^{*})^{2}} \right]\psi_{\rm far} =0.
\end{equation}
This is in the form of a Coulomb wave equation,
and can be solved with the Coulomb wave functions \cite{absteg}:
\begin{equation}
\label{eq:psifar2}
	\psi_{\rm far} = T\left(G_{1}(-\eta,\sqrt{\omega^{2}-\mu^{2}}\:r^{*})
	+ iF_{1}(-\eta,\sqrt{\omega^{2}-\mu^{2}}\:r^{*})\right)
\end{equation}
where
\begin{equation}
	\eta = \frac{\mu^{2} M}{\sqrt{\omega^{2}-\mu^{2}}}.
\end{equation}
When expanded near the interface with the region near the potential peak 
 we find that this
solution becomes (see Appendix \ref{sec:appc} for a detailed derivation):  
\begin{eqnarray}
	\psi_{\rm far} & \approx & \frac{iT \mu M^{1/2}}{3}  \frac{\pi^{1/2}e^{\pi \eta/2}}{\sqrt{\sinh(\pi \eta)}}
	(1+\eta^{2})^{1/2}(\omega^{2}-\mu^{2})^{3/4} 
	(r^{*})^{2} \nonumber \\
\label{eq:psifarapp2}
	 & & \qquad\qquad\qquad + \frac{T}{\mu M^{1/2}}\frac{\sqrt{\sinh(\pi \eta)}}
         {\pi^{1/2}e^{\pi \eta/2}}(1+\eta^{2})^{-1/2}
	 (\omega^{2}-\mu^{2})^{-1/4}\frac{1}{r^{*}}
\end{eqnarray}
This implies that we have replaced the $A$ and $B$ parameters of Equations (\ref{eq:A1})
and (\ref{eq:B1}) with the expressions:
\begin{equation}
	A = \frac{T}{\mu M^{1/2}}\frac{\sqrt{\sinh(\pi \eta)}}{\pi^{1/2}e^{\pi \eta/2}}(1+\eta^{2})^{-1/2}
         (\omega^{2}-\mu^{2})^{-1/4}
\end{equation}
and:
\begin{equation}
	B = iT\mu M^{1/2} \frac{\pi^{1/2}e^{\pi \eta/2}}{\sqrt{\sinh(\pi \eta)}}
        (1+\eta^{2})^{1/2}(\omega^{2}-\mu^{2})^{3/4}
\end{equation}
Combining these expressions with Equations (\ref{eq:A2}) and (\ref{eq:B2}), which still hold,
we can solve for $T$:
\begin{eqnarray}
	T &=& 2\Bigglb[\frac{2\mu M^{3/2}}{\omega}\frac{\pi^{1/2}e^{\pi \eta/2}}{\sqrt{\sinh(\pi \eta)}}
        (1+\eta^{2})^{1/2}(\omega^{2}-\mu^{2})^{3/4}+ \nonumber \\
	& & \qquad \qquad \qquad \frac{1}{2\mu M^{3/2}}
	\frac{\sqrt{\sinh(\pi \eta)}}{\pi^{1/2}e^{\pi \eta/2}}(1+\eta^{2})^{-1/2}
         (\omega^{2}-\mu^{2})^{-1/4}\Biggrb]^{-1}
\end{eqnarray}
and:
\begin{equation}
	\frac{1-R}{1+R} = \frac{4\mu^{2} M^{3}}{\omega}\frac{\pi e^{\pi \eta}}{\sinh(\pi \eta)}
	(1+\eta^{2})(\omega^{2}-\mu^{2}).
\end{equation}

These expressions for $T$ and $R$ show the distinction between a massless photon
and a massive vector boson in their explicit $\mu$ dependence. 
If $\mu \ne 0$, then the limit of vanishing $\omega^{2}-\mu^{2}$ gives:
\begin{equation}
	\lim_{\omega \rightarrow \mu} \left(\frac{1-R}{1+R}\right) = 8 M^{2} \pi \mu^{5}M^{5}.
\end{equation}
This equation tells us that even for the case of vanishing $\omega^{2}-\mu^{2}$ we can consistently define
three regions of solution for $\psi$ as long as $\mu \ne 0$.  This fact
allows us to construct transmission and
reflection coefficients for the barrier.  Therefore, we can be confident that the
transmission of massive vector waves really does approach zero as $\omega$ becomes
small.  There are no technicalities to save an external massive vector field.

Now that we have decided that massive vector waves can be said to reflect from the
curvature
barrier, we must decide what that statement means.  In particular, we need to 
understand the timescales for reflection and transmission.  It is
important to note that our construction of the transmission and reflection coefficients
does not allow us to pinpoint the location where reflection takes place.  Rather,
we must consider the reflection to be a result of the accumulated effect of the
potential barrier over its entire width.  We can say with certainty, however, that
it is the edge of the barrier that is farthest from
the black hole which plays the decisive role.  If we can consistently define such
an edge as the place where the terms generated by the mass of the Proca field begin
to dominate the
curvature potential, we expect reflection and loss of the field.  If we cannot
define such an edge, then a static $1/r^{2}$ law holds and we expect to find a
Reissner-Nordstr\"om black hole.  

We can estimate the approximate location of this edge by comparing the magnitude
of the terms in the potential of the far zone.  It is apparent from Equation
(\ref{eq:fieldpsi}) that 
the changeover from a curvature dominated potential to
a mass-dominated potential will happen when:
\begin{equation}
	\frac{2}{r^{2}} \sim \mu^{2}
\end{equation}
which gives a characteristic location:
\begin{equation}
	r \approx \mu^{-1}.
\end{equation}

To approximate the time taken for waves to begin reflecting, we can construct
two separate arguments leading to the same basic result.  The clearest argument
is that unless $\omega$ is smaller than $\mu$, the $\mu$ term is negligible in
the field equation.  Thus, our first inclination that the field is not massless
must come from waves with a characteristic timescale $\mu^{-1}$.  Therefore,
we expect the external field to persist (as in the massless case) until a time
of order $\mu^{-1}$ has elapsed.  Another way to arrive at this conclusion is to 
consider what we have already said about the mechanics of reflection.  Since
it is the far edge of the potential at $r \sim \mu^{-1}$ which will make the
final
difference between reflection and transmission, an outgoing wave cannot 
in effect be reflected until it has time to feel the effects of the \emph{entire}
potential barrier, all the way out to $r \sim \mu^{-1}$.  Assuming then that
the wave travels near the speed of light (this will at least be good for a lower
bound) the time that must elapse between the onset of collapse and the
first reflection of outgoing waves originating near the collapsing surface is of order $\mu^{-1}$.

\subsection{Numerical Solution}

The previous section has motivated the idea that an external massive gauge field
arising from sources that collapse into a black hole will 
decay with
a characteristic timescale set by the mass of the gauge field.  Now, we
wish to provide a more concrete demonstration.
To do so, we can numerically integrate the field equation for $\Phi_{0}$.

The first step in this process is to establish the initial conditions
for the integration.  We will follow Price \cite{Prisc} in
considering the case of a static star with initial radius $R_{i} \sim 4M$
which suddenly begins to collapse into a black hole of mass $M$.
This situation implies that at the moment when collapse begins, the
field far away from the star should resemble the flat space static 
Proca solution.
Thus, we expect that the ``electric field'' far from the star will have
the form:
\begin{equation}
	E = Q e^{-\mu r}\left[\frac{\mu}{r} + \frac{1}{r^{2}}\right]
\end{equation}
where $Q$ is the total net Proca charge (in some normalized units) contained
within the star.  Since we know that $\Phi_{0}$ is equal to the ``electric field''
for the Schwarzschild space Proca solution (see the discussion in Appendix \ref{sec:appb} and around Equation 
(\ref{eq:eform})), we can use the flat space form
for $E$ to set the field value and first derivative of $\Phi_{0}$ at some
point far from the initial surface of the star.  Then, by numerically integrating
the field equation for $\Phi_{0}$ outside the star
(Equation (\ref{eq:fieldphi})),
we can find the initial conditions for $\Phi_{0}$ everywhere outside the star
before collapse.  

To make this integration simpler, we will assume that our end result will resemble
the flat space Proca solution everywhere (this assumption will be justified in our results).
With this assumption, it becomes useful to make the substitution:
\begin{equation}
\label{eq:numsub}
\Phi_{0}(r,t_{i}) \equiv f(r) e^{-\mu r}\left(\frac{1}{r^{2}}+\frac{\mu}{r}\right)
\end{equation}
which results in the field equation for $f(r)$:
\begin{eqnarray}
\frac{2M-r}{r^{3}(1+\mu r)} 
 &[- \left(1-\frac{2M}{r}\right) (1 + \mu r)r^2
f''(r)
+ 2(r(1+\mu r + \mu^{2} r^{2}) - & \nonumber \\ 
\label{eq:numred}
& M(3 + 3\mu r +
2 \mu^{2} r^{2})) f'(r)
+2M\mu^{2} r \left(
2+\mu r\right) f(r)]
& = 0. 
\end{eqnarray}

\begin{figure}
\includegraphics[width=\columnwidth]{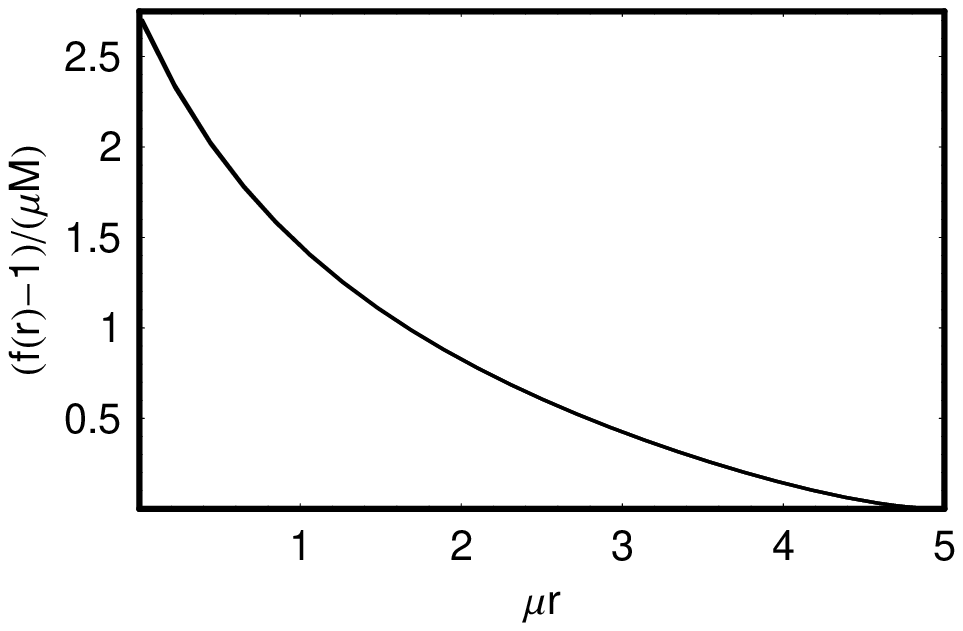}
\caption{Results of integration for $f(r)$ of Equation \ref{eq:numred}.
This figure shows the results for $\mu=0.01$ and $\mu=0.001$ overlaid 
(indistinguishable).}
\label{fig:phiinit}
\end{figure}

\begin{figure}
\includegraphics[width=\columnwidth]{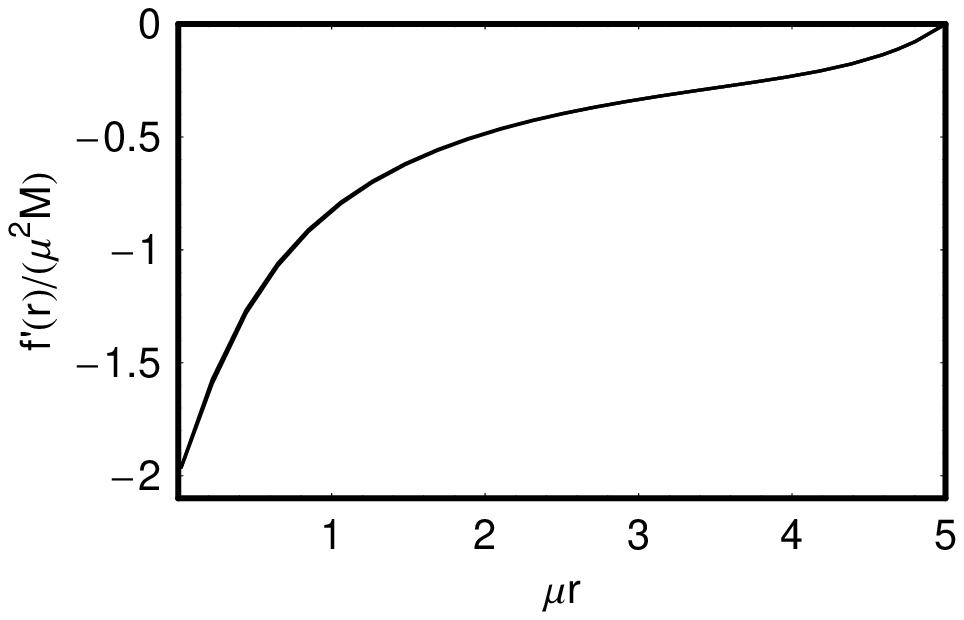}
\caption{Results of integration for $f'(r)$ of Equation \ref{eq:numred}.
This figure shows the results for $\mu=0.01$ and $\mu=0.001$ overlaid
(indistinguishable).}
\label{fig:phipinit}
\end{figure}

\begin{figure}
\includegraphics[width=\columnwidth]{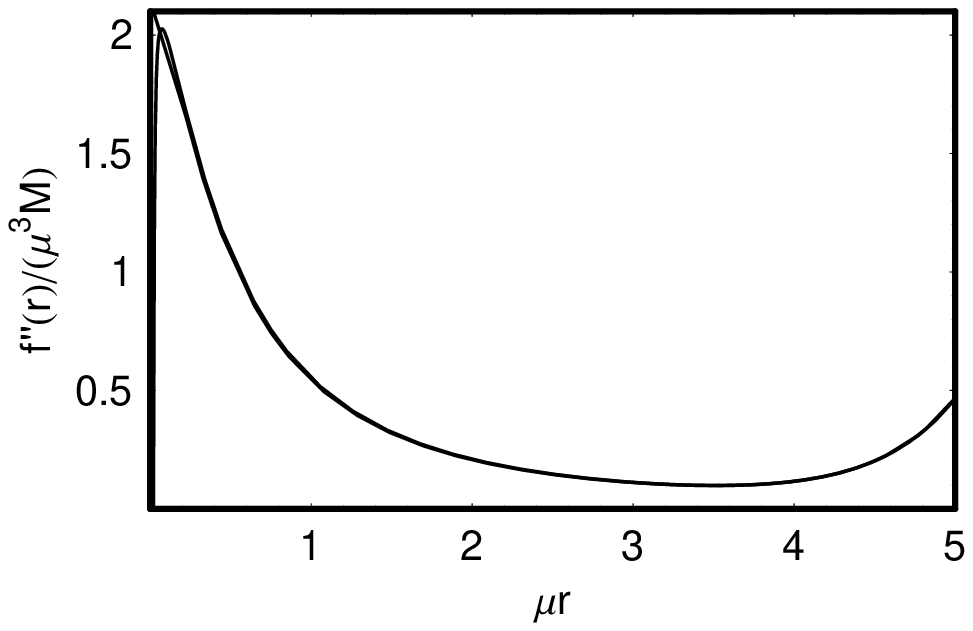}
\caption{Results of integration for $f''(r)$ of Equation \ref{eq:numred}.
This figure shows the results for $\mu=0.01$ and $\mu=0.001$ overlaid
(indistinguishable).}
\label{fig:phippinit}
\end{figure}

In terms of $f(r)$, the boundary conditions for our numerical integration
are particularly simple.  Far from the location
of the star ($r \gg M$) we expect $f = 1$ and $f' = 0$.  We choose to take
$f(5/\mu) = 1; \; f'(5/\mu) = 0$ as our far boundary condition and integrate
in toward the initial radius of the star.
The results of this integration are summarized in Figures \ref{fig:phiinit}--\ref{fig:phippinit}.

From these results, one can see that the initial conditions for $\Phi_{0}$
just outside the star will deviate from the Proca solution
at the order $\mu M$.
Now we must consider
how the field evolves in the area vacated by the collapsing star once 
collapse to a black hole has begun (the region between $R_{i}$ and
the eventual horizon $2M$).  Here we must appeal to an analogy with electromagnetism.

In the case of a zero mass field, $\Phi_{0}^{\rm EM}$ would be $Q/r^{2}$ 
everywhere outside the initial surface of the star.   In this case,
the Reissner-Nordstr\"om solution tells us that upon collapse of the star
the $\Phi_{0}^{\rm EM}$ field will simply evolve to equal $Q/r^{2}$ everywhere 
external to the black hole.  
This is important, because numerical integrations for the Proca field show that if
$\mu^{2}R_{i}^{2} \sim \mu^{2}M^{2} << 1$ we
obtain initial conditions 
that closely resemble those for a massless EM field.
In fact, our initial conditions on $\Phi_{0}$ 
near the surface of the star can be summarized:
\begin{eqnarray}
	\Phi_{0}(Q,R_{i},t_{i}) & \equiv & \Phi_{0}^{\rm EM}(\tilde{Q},R_{i},t_{i}) \\
	\partial_{r} \Phi_{0}(Q,R_{i},t_{i}) &=& \partial_{r}\Phi_{0}^{\rm EM}(\tilde{Q},R_{i},t_{i}) \left(1+\mathcal{O}(\mu^{2}R_{i}^{2})\right) \\
	\partial^{2}_{r} \Phi_{0}(Q,R_{i},t_{i}) & =&  \partial^{2}_{r}\Phi_{0}^{\rm EM}(\tilde{Q},R_{i},t_{i}) \left(1+\mathcal{O}(\mu^{2}R_{i}^{2})\right) 
\end{eqnarray}
where $\tilde{Q} = Q\left(1+\mathcal{O}(\mu R_{i})\right)$ renormalizes the EM 
profile (this is allowed because the overall normalization is irrelevant to the evolution
of the EM field -- we would obtain a $Q/r^{2}$ profile regardless of the
value of $Q$).
If we now take $R_{i}$ of order $4M$ then we see from
Figure \ref{fig:procapot} that
at radial distances $r < R_{i}$ the effective potential governing
the evolution of $\Phi_{0}$ rapidly approaches that for electromagnetism (differing by
less than order $\mu^{2}M^{2}$).
Putting these initial conditions and the effective potential into the differential equation
governing the evolution of the field (Equation (\ref{eq:fieldphi})) shows us that the
evolution of the $\Phi_{0}$ field 
in the region vacated by the collapsing star will be governed by a differential equation that
differs from that for a massless EM field only at the order $\mu^{2} M^2$.  Further, these
differences will be suppressed by the factor $(1-2M/r)$ which rapidly becomes small in 
this region.  Similarly, the integration of the field equation will use boundary conditions
that differ from the EM case only at the order $\mu^{2} M^{2}$ and the collapse will characteristically
require a time $M \ll \mu^{-1}$.
Taken together, all of this implies that we expect the profile of a massive vector monopole
field
to differ by $\mathcal{O}(\mu^{2} M^{2})$ from the EM case in the near neighborhood of
the event horizon ($r \sim M$) immediately after the star
has completed its collapse to a black hole. 

Our analogy with the EM case has shown us that the Proca field in the region vacated by
the collapsing star will have a profile that is within order $\mu^{2} M^{2}$ of the
EM field result $\tilde{Q}/r^{2}$ immediately following the collapse. 
This, in turn, differs only at the order $\mu M$ from
the flat space Proca profile.
We have already shown through explicit numerical integration that the field
outside the original surface of the star
will also be within order $\mu M$ of the flat space Proca profile.
We can therefore assume that up to errors of order $\mu M \ll 1$:
\begin{equation}
	\Phi_{0}(t=0) = Q \exp\left(-\mu r\right)\left[\frac{\mu}{r} + \frac{1}{r^{2}}\right]
\end{equation}
describes the Proca field profile \emph{everywhere} outside the horizon of the
black hole immediately after collapse.  Since we are trying to show that the field will persist
well after the black hole has collapsed, we will begin our simulations
after the black hole has already formed.

\begin{figure*}
\begin{center}
\includegraphics{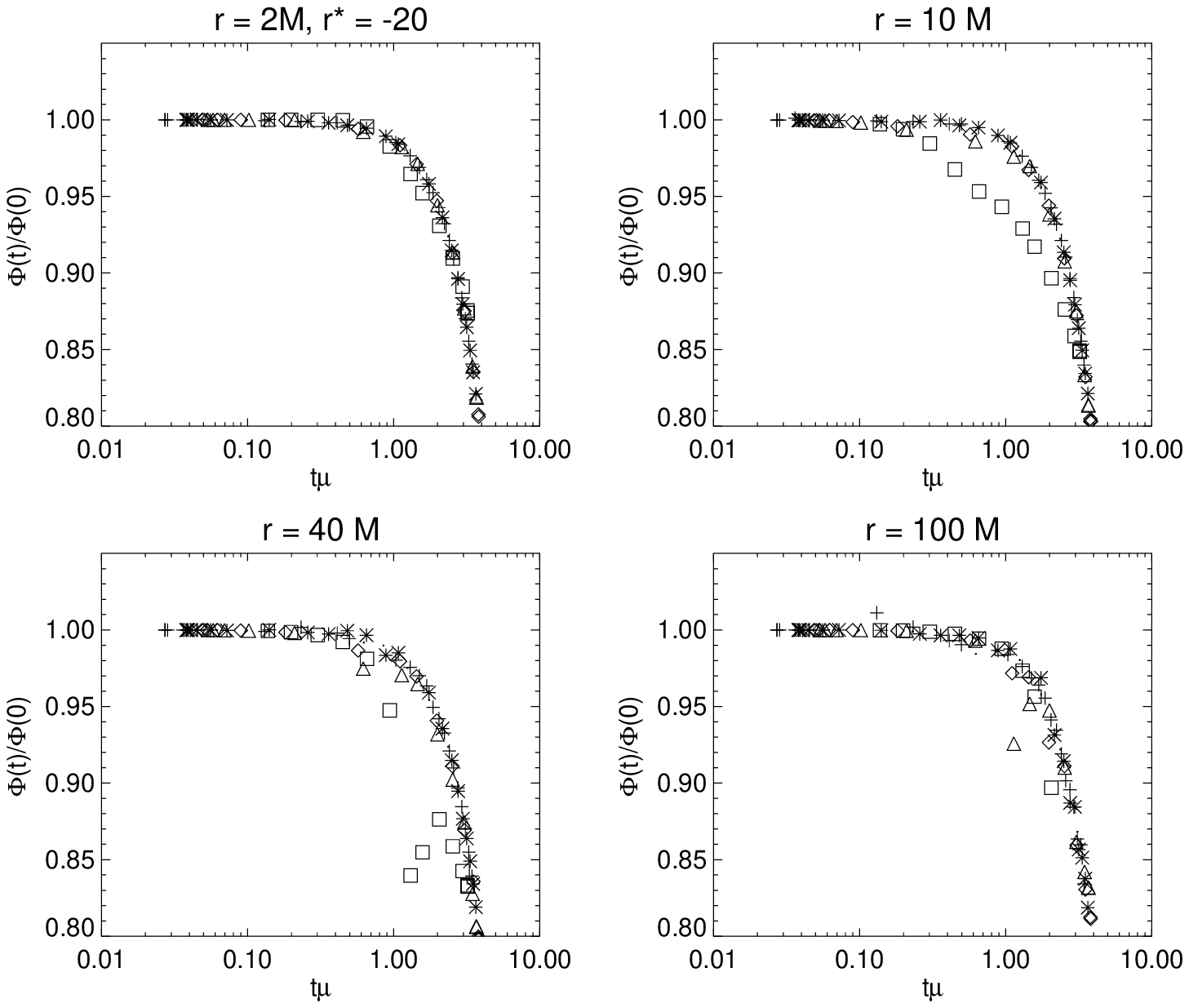}
\caption{Evolution of the $\Phi$ field for various values of $\mu M$:
 0.001 (+); 0.0025 ($\ast$); 0.005 ($\cdot$); 
0.0075 ($\Diamond$); 0.01 ($\triangle$); 0.025 ($\Box$).}
\label{fig:phidecay}
\end{center}
\end{figure*}

\begin{figure*}
\includegraphics{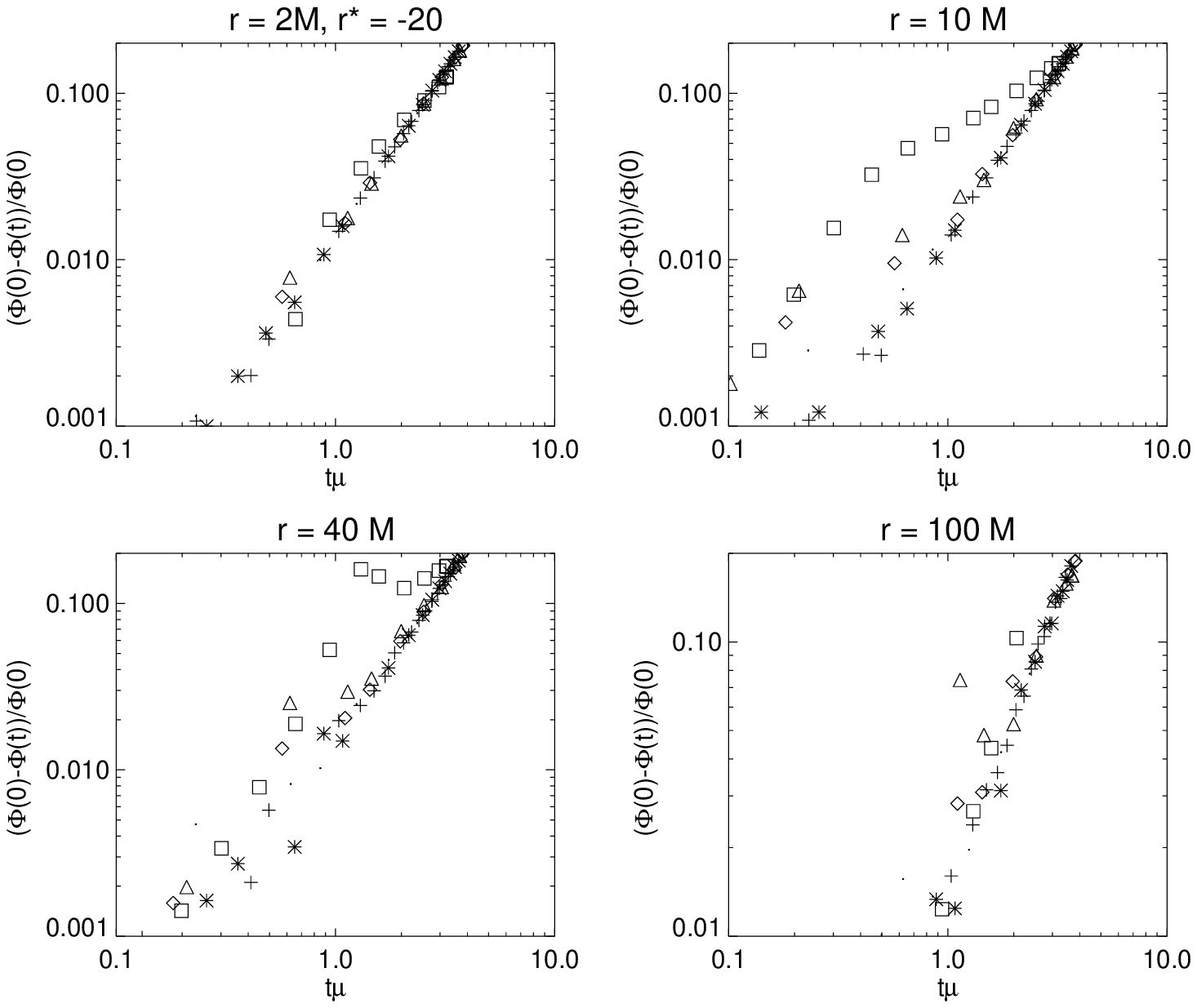}
\caption{Log plot of the evolution of $\Phi_{0} - \Phi(t)$ for various
values of $\mu M$:
 0.001 (+); 0.0025 ($\ast$); 0.005 ($\cdot$);
0.0075 ($\Diamond$); 0.01 ($\triangle$); 0.025 ($\Box$).} 
\label{fig:philog}
\end{figure*}

Using the static Proca field solution as our initial conditions, we
numerically integrate the full field equation (\ref{eq:fieldphi}) for several
values of $\mu M$.  The results of these integrations are summarized in Figures
\ref{fig:phidecay} and \ref{fig:philog}.  These figures clearly show that the
timescale for loss of the field is characteristically of order $\mu^{-1}$,
and that the
field near the black hole remains almost completely undisturbed for a time
approximately equal to $\mu^{-1}$ as long as $\mu M < 1/40$.  
For $\mu M > 1/40$ we can make no definite statement about the field behavior, since in
this regime the
field evolution is sufficiently violent that a more sophisticated numerical integration
technique is needed (also, one should begin to consider evolution of the field during the
collapse of the star).  

\section{Consequences}

\label{sec:consequences}

In this section we wish to discuss possible
values of $\mu M$.  We therefore switch from the natural units
of general relativity ($G= \hbar = c = 1$) to the natural units of
field theory ($\hbar = c = 1$, $G = 1/M_{pl}^{2}$).  The major difference
is that $M$ no longer has the units of radius.  Instead, the Schwarzschild
radius is $2M/M_{pl}^{2}$.  Thus, rather than
the condition $\mu M \ll 1$, the correct expression in field theory
units is: $\mu M/M_{pl}^{2} \ll 1$.

Once we have shown that a massive vector field around a star collapsing to
a black hole will persist for a time of order $\mu^{-1}$, we can draw an
important conclusion.  The lifetime of a black hole goes as $M^{3}$ \cite{hawkrad}.
Since a Planck-mass black hole is expected to live one Planck time,
we have the approximate relation: 
\begin{equation}
	t_{evap} \sim \frac{M^{3}}{M_{pl}^{4}}.
\end{equation}
This approximation is essentially borne out by the calculations of Page \cite{Page1},
up to 
a numerical factor which
depends on the particle species that are emitted by the black hole.  For
the smallest black holes, one must include many known species not accounted for
in Page's work and perhaps some wholly unknown species.  These extra
particles will tend to lessen the lifetime of small black holes, bringing
the results of \cite{Page1} closer to our own approximation.

As long as $t_{evap}$ is less than $\mu^{-1}$, we anticipate that the external
field will persist for the \emph{entire life} of the black hole (unless,
of course, the charge is recovered before the complete evaporation of the
black hole).  If the black
hole can evaporate while the external field persists, we anticipate the recovery
of the charge \cite{Gibbons,Carter,Page}.  We can therefore estimate the black
hole mass below which a symmetry broken at the scale $\mu$ will obey charge
conservation.  The result is:
\begin{equation}
	M < \frac{M_{pl}^{4/3}}{\mu^{1/3}}.
\end{equation}
In a traditional universe with the Planck scale at about $10^{19}$~GeV, this
implies that a symmetry broken at a scale of $10^{4}$ GeV (just beyond
the reach of current colliders) would be respected
by black holes up to $10^{5}$ Planck masses (approximately 1 g).  

\section{Extrapolations}

It has historically been assumed that virtual or quantum black holes
would not respect a broken symmetry because classical black holes do not.  Here
we have shown, however, that there is a strong case to be made for the
claim that the smallest classical black holes will respect a broken symmetry.  What constitutes
a ``small'' black hole depends on the scale at which the symmetry is broken. 
Symmetries broken at small scales (high energies) require less massive (shorter-lived)
black
holes than symmetries broken at low energy.  Since Planck-scale
black holes are smaller and shorter-lived than any classical black hole,
this result calls into question the conclusion that broken symmetries are
not respected by quantum gravity.

This result could be of particular importance if there is a low scale for quantum
gravity.  Here, virtual black holes have been predicted to result in fast
proton decay \cite{Gordy}.  Now, however, we see that protons might
be protected by a broken gauge symmetry.  In fact, it is conceivable
that the scale for such symmetry breaking can be pushed nearly to the
quantum gravity scale, leaving the possibility that it decouples
from physics at currently explored energies even in a low Planck
scale universe.

\appendix

\section{Derivation of the Field Equations}

\label{sec:appa}

We begin our derivation of the field equations for a massive vector
field in Schwarzschild spacetime
 by following Price in making use of the spinor formalism to 
write the field equation for a massive gauge boson in terms of scalar
quantities.  To obtain the correct expression for the field tensor
$F_{\mu\nu}$ it is necessary to supplement the expressions given in
Price's work \cite{Price} with the formalism found in the article
by Pirani \cite{Pirani}.  In this way we arrive at the expression:
\begin{equation}
	F^{\mu \nu} = \sigma^{\mu}_{A\dot{X}}\sigma^{\nu}_{B\dot{Y}}
	F^{A\dot{X}B\dot{Y}}
\end{equation}
where Latin capitals denote spinor indices (see \cite{Pirani} for more
detail).

We can now choose the connection $\sigma^{\mu}_{A\dot{X}}$ to
be determined by the null tetrad chosen to parameterize our spacetime \cite{NP}.
We follow Price in using the null tetrad:
\begin{eqnarray}
	 l^{\mu}& =& \left(\left(1-\frac{2M}{r}\right)^{-1},1,0,0\right) \\
	 n^{\mu}& =& \frac{1}{2}\left(1,-\left(1-\frac{2M}{r}\right),0,0\right) \\
	 m^{\mu} &=& \frac{1}{\sqrt{2}}\left(0,0,\frac{1}{r},\frac{i}{r \sin\theta}\right) 
\end{eqnarray}
We then choose:
\begin{equation}
	\sigma^{\mu}_{00}=l^{\mu}; \:\: \sigma^{\mu}_{11}=n^{\mu}; \:\:
	\sigma^{\mu}_{01}=m^{\mu}; \:\:  \sigma^{\mu}_{10}=\overline{m}^{\mu}
\end{equation}
where $\overline{m}^{\mu}$ denotes the complex conjugate of $m^{\mu}$.
The conversion to scalar fields is now made via the definition \cite{Price}:
\begin{equation}
	F_{A\dot{X}B\dot{Y}}=\frac{1}{2}\left(\epsilon_{AB}\Phi^{*}_{\dot{X}\dot{Y}}
	+ \epsilon_{\dot{X}\dot{Y}}\Phi_{AB}\right).
\end{equation}

Using the rules for contracting spinor indices (see \cite{Pirani}), we can 
express the field tensor in the form:
\begin{eqnarray}
F^{\mu \nu} & = & 
 	 \frac{1}{2}[- l^{\mu}n^{\nu}(\Phi_{10}+\Phi^{*}_{10}) 
		+ l^{\nu}n^{\mu}(\Phi_{01}+\Phi^{*}_{01})  
	 + m^{\mu}\overline{m}^{\nu}(\Phi_{10}-\Phi^{*}_{01}) +\; c.c.\; \nonumber \\
\label{eq:Fphi}
	& &   \mbox{} + (l^{\mu}m^{\nu}-l^{\nu}m^{\mu})\Phi_{11}+\; c.c.\; 
	+(m^{\mu}n^{\nu}-m^{\nu}n^{\mu})\Phi^{*}_{00}+\;c.c.] 
\end{eqnarray}
where $c.c.$ denotes the complex conjugate of the term immediately preceding.
Note that this expression for $F^{\mu \nu}$
is manifestly antisymmetric and real as long as $\Phi_{10}=\Phi_{01}$,
which is guaranteed by the total symmetry of spin indices \cite{Price,Pirani}.
We desire an antisymmetric, real $F^{\mu \nu}$ so that we can make the
standard
definition:
\begin{equation}
\label{eq:trad}
	F_{\mu \nu} = \partial_{\mu} A_{\nu} - \partial_{\nu}A_{\mu}.
\end{equation}
We henceforth simplify our notation by introducing:
 $\Phi_{0}\equiv\Phi_{10}$, $\Phi_{+1}\equiv\Phi_{00}$, and $\Phi_{-1}\equiv\Phi_{11}$.

We now have the tools necessary to derive the field equations for the Proca field.
To obtain a convenient form, there is one more trick to use.
We combine the homogeneous and inhomogeneous Proca equations (Maxwell equations
for a massive vector field) into the form \cite{Pirani}:
\begin{equation}
	F^{\mu \nu}_{\ \ ;\nu} + \frac{i}{2\sqrt{-g}}(\epsilon^{\alpha\beta\mu\nu}
	F_{\alpha \beta})_{;\nu} = \mu^{2} A^{\mu}.
\end{equation}
This yields the equations (assuming $\Phi_{0}$ has
the form of a spherical harmonic):
\begin{eqnarray}
	\label{eq:fld1}
	D\left(r^{2}\hat{\Phi}_{0}\right) & = & r\hat{\Phi}_{+1} + \mu^{2}r^{2}
	\left[A^{0}-S^{-1}(r) A^{1}\right] \\
	\label{eq:fld2}
	2\Delta \left(r^{2}\hat{\Phi}_{0}\right) & = &  2r\hat{\Phi}_{-1}
	 - r^{2}\mu^{2}\left[S(r) A^{0}+A^{1}\right] \\
 	2D\left(r\hat{\Phi}_{-1}\right)& = & - l(l+1)\hat{\Phi}_{0}	
	+ \mu^{2}r^{2}
	\: \Theta \! \left[i\sin\theta A^{3}-A^{2}\right] \\
	\label{eq:fld4}
	2\Delta\left(r S(r)\hat{\Phi}_{+1}\right) & = & - l(l+1)
	S(r)\hat{\Phi}_{0}
	+ \mu^{2}r^{2} S(r) \:
        \Theta \! \left[A^{2}+i\sin\theta A^{3}\right].
\end{eqnarray}
Here we have defined a number of quantities to simplify the form of the final equations.  
They are:	
\begin{eqnarray}
	D & \equiv &  \left(1-\frac{2M}{r}\right)^{-1}\left[\partial_{t}+\partial_{r^{*}}\right] \\
	\Delta &\equiv & \frac{1}{2}\left[\partial_{t}-\partial_{r^{*}}\right] \\
	\Theta & \equiv &\left(\frac{1}{\sin\theta} \partial_{\theta} \sin\theta
        +\frac{i}{\sin\theta}\partial_{\phi}\right)	\\
	S(r)&\equiv &\left(1-\frac{2M}{r}\right). 
\end{eqnarray}
Further, we have employed the ``despun'' versions of the fields as defined in \cite{Price}:
\begin{eqnarray}
	\hat{\Phi}_{0} &\equiv & \Phi_{0} \\
	\hat{\Phi}_{+1} & \equiv & \frac{1}{\sqrt{2}}\left(\partial_{\theta}
	- \frac{i}{\sin\theta}\partial_{\phi} + \cot\theta\right) \Phi_{+1} \\
	\hat{\Phi}_{-1} & \equiv & \frac{1}{\sqrt{2}}\left(\partial_{\theta}
	+ \frac{i}{\sin\theta}\partial_{\phi} + \cot\theta\right)\Phi_{-1}.
\end{eqnarray}
One benefit of using these definitions is that it is immediately obvious that
the equations for a massive vector boson field reduce to Price's equations for
a massless vector boson field (given as Equation (42) in \cite{Price}) when we
take the limit $\mu \rightarrow 0$.

\section{Monopole Fields}

\label{sec:appb}

So far we have not assumed a monopole field.  In fact, the results of
Appendix \ref{sec:appa}
are general in the sense that any field can be decomposed into spherical
harmonics.  In this Appendix we will specialize to the case of spherical symmetry.
The monopole produces the obvious simplification that $l=0$.  
There are also restrictions on the form of $F^{\mu \nu}$ imposed by
spherical symmetry (assuming $F^{\mu \nu}$ is given in terms of
potentials as in equation (\ref{eq:trad})).  Because the $\Phi$ fields
are defined in terms of $F^{\mu \nu}$ (see Equation (\ref{eq:Fphi})),
spherical symmetry guarantees:
\begin{eqnarray}
	\Phi_{0} & = & \Phi_{0}^{*} \\
	\Phi_{+1}& = & \Phi_{-1} = 0.
\end{eqnarray}
Further, by comparing Equations (\ref{eq:Fphi}) and (\ref{eq:trad}),
we can see that in the spherically symmetric case:
\begin{equation}
	F_{10} = \Phi_{0}.
\end{equation}
By analogy with the case of electromagnetism, we see that $\Phi_{0}$
parameterizes the effective ``electric field'' of the broken symmetry.

Using these relationships, we can combine Equations (\ref{eq:fld1}) and
(\ref{eq:fld2}) into a single, second order differential equation.  
The resulting field equation is exactly Equation (\ref{eq:fieldphi}).

\section{Coulomb Wave Functions}

\label{sec:appc}

Equation (\ref{eq:omegeqmu}) is solved by the Coulomb wave functions
$F_{1}(-\eta,\rho)$ and $G_{1}(-\eta,\rho)$, where in
our case \cite{absteg}:
\begin{equation}
	\eta = \frac{\mu^{2}M}{\sqrt{\omega^{2}-\mu^{2}}}
\end{equation}
and:
\begin{equation}
	\rho = \sqrt{\omega^{2} - \mu^{2}} \; r^{*}.
\end{equation}

To consistently define a transmitted wave, we wish to find the linear combination
of $F_{1}$ and $G_{1}$ which asymptotes to $e^{ikr^{*}}$ as $r^{*}$ (and hence $\rho$)
approaches infinity.
For $\rho \rightarrow \infty$, we find \cite{absteg}:
\begin{equation}
	\lim_{\rho \rightarrow \infty} F_{1}(-\eta,\rho) = \sin(\rho+\phi)
\end{equation}
where $\phi$ is a phase angle.
Similarly, in the same limit:
\begin{equation}
	\lim_{\rho \rightarrow \infty} G_{1}(-\eta,\rho) = \cos(\rho+\phi).
\end{equation}
Thus, our transmitted wave will have the form:
\begin{equation}
	\psi_{\rm far} = T(G_{1}(-\eta,\rho)+iF_{1}(-\eta,\rho))
\end{equation}
just as asserted in Equation (\ref{eq:psifar2}).

Now, we wish to look at the expansion of $\psi_{\rm far}$ in the limit of small
$\rho$, which will be matched to the solution for $\psi$ near the peak of the
curvature potential $(r^{*} \sim few \times M)$.  For $\rho \ll 1$, we
have \cite{absteg}:
\begin{equation}
	F_{1}(-\eta,\rho) \approx C_{1}(-\eta) \rho^{2}
\end{equation}
where:
\begin{equation}
	C_{1}(-\eta) = \frac{\exp(\pi \eta/2) |\Gamma(2-i\eta)|}{3}.
\end{equation}
We can express $|\Gamma(2-i\eta)|$ in the form \cite{absteg}:
\begin{equation}
	|\Gamma(2-i\eta)|^{2} = (1+\eta^{2})\frac{\pi \eta}{\sinh(\pi \eta)}
\end{equation}	
which gives:
\begin{equation}
       F_{1}(-\eta,\rho) \approx \frac{\sqrt{\pi}}{3}\frac{e^{\pi \eta/2}}{\sqrt{\sinh(\pi \eta)}}
        \eta^{1/2}(1+\eta^{2})^{1/2}\rho^{2}.
\end{equation}
Similarly, for $\rho \ll 1$, we can expand $G_{1}$ to find \cite{absteg}:
\begin{equation}
	G_{1}(-\eta,\rho) \approx \frac{1}{3C_{1}(-\eta)\rho}
\end{equation}
where $C_{1}$ is the same function as above, giving:
\begin{equation}
	G_{1}(-\eta,\rho) \approx \frac{1}{\sqrt{\pi}}\frac{\sqrt{\sinh(\pi \eta)}}
	{e^{\pi \eta/2}}\eta^{-1/2}(1+\eta^{2})^{-1/2}
         \frac{1}{\rho}.
\end{equation}
These limiting forms give us exactly Equation (\ref{eq:psifarapp2}).

\begin{acknowledgments}

I wish to thank P. Davies for a helpful discussion and encouragement.  I also thank the
referee for pointing out a mistake in the original draft.

\end{acknowledgments}

\end{document}